\documentclass[12pt]{article}
\usepackage{amssymb,amsmath,epsfig}
\allowdisplaybreaks

\begin{document}
\title{\bf Traversable Wormhole Solutions admitting Karmarkar Condition in $f(R,T)$ Theory}
\author{M. Sharif \thanks {msharif.math@pu.edu.pk} and Arooj Fatima
\thanks{arooj3740@gmail.com}\\
Department of Mathematics and Statistics,\\
The University of Lahore, 1-KM Defence Road Lahore, Pakistan.}
\date{}
\maketitle

\begin{abstract}
In this paper, we evaluate traversable wormhole solutions through
Karmarkar condition in $f(R,T)$ theory, where $T$ is the trace of
the energy-momentum tensor and $R$ represents the Ricci scalar. We
develop a wormhole shape function for the static traversable
wormhole geometry by using the embedding class-I technique. The
resulting shape function is used to construct wormhole geometry that
fulfills all the necessary conditions and joins the two
asymptotically flat regions of the spacetime. We investigate the
existence of viable traversable wormhole solutions for anisotropic
matter configuration and examine the stable state of these solutions
for different $f(R,T)$ gravity models. We analyze the graphical
behavior of null energy bound to examine the presence of physically
viable wormhole geometry. It is found that viable and stable
traversable wormhole solutions exist in this modified theory of
gravity.
\end{abstract}

\section{Introduction}

Our universe raises marvelous questions for the scientific community
due to its mysterious nature. The existence of hypothetical
structures are considered the most controversial issues that create
the wormhole (WH) structures. It is defined as a speculative tunnel
through which observers may be able to travel freely from one part
of spacetime to another. An intra-universe WH is a bridge that
connects distant parts of the same spacetime, while an
inter-universe WH joins distant regions of the separate spacetime.
In 1916, Flamm \cite{1} established WH geometry through the
Schwarzschild solution. Later, Einstein and Rosen \cite{2}
demonstrated that a curved-space structure can connect two separate
parts of the universe through a tunnel named as Einstein-Rosen
bridge. In 1950, Wheeler \cite{3} found that Schwarzschild WH is
non-traversable because two-way traveling is not possible in it, and
anything attempting to pass through would be destroyed by the
tremendous tidal forces present at the WH throat. Moreover, the WH
throat rapidly expands from zero to finite circumference and then
compresses to zero with time, thus prevents the access to anything.
However, Wheeler and Fuller \cite{4} investigated that WHs would
collapse instantly after the formation.

The existence of a viable WH geometry is questioned due to a large
amount of exotic matter (which violates energy conditions). Hence,
the exotic matter in the bridge must be minimum for a viable WH
structure. The first traversable WH was proposed by Morris and
Thorne \cite{5}. One may confine the exotic matter at the WH throat
with the help of matching conditions \cite{6}. Vacaru et al \cite{8}
studied the anisotropic WH solutions through a holonomic approach.
Wormholes can be classified into static and dynamic WHs \cite{9}. In
contrast to non-static WHs configuration, static WHs have a fixed
radius on the throat. Dzhunushaliev et al \cite{10} investigated the
stability of WH configurations in the presence and absence of an
electromagnetic field. The static spherically symmetric Lorentzian
WH solutions have been explored in \cite{11}. The study of WH shape
function with its properties is one of the interesting subject in
traversable WH geometry. Recently, some people \cite{12}-\cite{14}
proposed different shape functions to explain the WH geometry.

Several methods have been used to analyze the WH geometry such as
particular form of the equation of state, the solution of metric
potentials and constraint on matter variables. In this regard, the
embedding class-I technique has been established that provides a
relation between radial and time coordinates which helps to study
the cosmic objects. According to this strategy, an $n$-dimensional
manifold can be embedded into $(n+m)$-dimensional manifold. The
embedding class-I condition is developed to examine static
spherically symmetric solution \cite{15}. A necessary condition for
a static spherically symmetric line element which belongs to class-I
was developed by Karmarkar \cite{16}. Recently, spherically
symmetric objects with various matter configurations through the
Karmarkar condition have been discussed \cite{17}-\cite{21}. Fayyaz
and Shamir \cite{22} used Karmarkar condition to examine the viable
as well as stable WH structures.

In the literature, several hypotheses have been proposed to analyze
dark energy such as quintessence fields, positive cosmological
constant and Chaplygin gas. To explain the accelerated expansion of
the universe, one may only need to include the cosmological constant
or any other exotic resource. In this perspective, modified
gravitational theories are considered the most effective techniques
to describe the cosmic mysteries. Accordingly, one of the simplest
modified theory is $f(R)$ gravity which describes dark energy and
accelerated expansion of the universe. Harko et al \cite{23}
generalized $f(R)$ theory by introducing the concept of
curvature-matter coupling and named as $f(R,T)$ theory of gravity.
This modified theory is non-conserved that ensures the presence of
an extra force and consequently, non-geodesic motion of particles.
Houndjo \cite{24} numerically established holographic dark energy
model to discuss the cosmic expansion and found that this theory
reproduces a similar extension history as produced by general
relativity (GR). In the $f(R,T)$ framework, many cosmic topics have
been studied such as anisotropic solutions, energy constraints,
thermodynamics, viscosity solution, phase space and stability
analysis \cite{25}-\cite{39a}.

The physical viable features of WHs yield interesting results in the
context of alternative theories of gravity. Lobo and Oliveira
\cite{41} studied traversable WH geometry through different shape
functions and equations of state in $f(R)$ gravity. Bertolami and
Ferreira \cite{42} examined that curvature-matter coupled theory
yields viable WH solutions. The viable traversable WH geometry with
a specific equation of state in $f(R,T)$ gravity has been examined
in \cite{43}. Sharif and Fatima \cite{44} considered viable $f(G)$
model to investigate static spherically symmetric WH solutions (G is
the Gauss-Bonnet invariant). Sharif and Shahid \cite{45} explored
viable static WH solutions through Noether symmetry technique in the
framework of $f(G,T)$ gravity. Shamir et al \cite{46} considered
static spherical geometry with anisotropic fluid source to discuss
WH solutions in $f(R)$ theory. Mustafa at al \cite{47} obtained
physically realistic traversable WH solutions through the Karmarkar
condition in $f(Q)$ gravity (Q is the non-metricity scalar).

Shamir and Fayyaz \cite{48} constructed a shape function through
Karmarkar condition in $f(R)$ gravity and found that WH structure
can be obtained with a negligible amount of exotic matter. Sharif
and Gul \cite{49} examined static WH solutions through Noether
symmetry approach in the context of $f(R,T^2)$ theory, where $T^2$
is the self-contraction of energy-momentum tensor. Naz et al
\cite{49a} investigated the geometry of compact stars by using
Karmarkar condition in the background of $f(R)$ gravity. Mustafa et
al \cite{49b} examined the Gaussian and Lorentzian distributed
spherically symmetric WH solutions in $f(\tau,T)$ gravity, where
$\tau$ is torsion. Godani \cite{49c} explored the traversable WHs in
the context of $f(R,T)$ theory of gravity with $f(R,T)=R+2\lambda
T$, where $\lambda$ is an arbitrary constant. Malik et al \cite{50}
examined traversable WH models in $f(R)$ theory by applying the
Karmarkar condition.

In this paper, we investigate traversable WH solutions in $f(R,T)$
gravity by using the Karmarkar condition. For this purpose, we
evaluate the behavior of the shape function and null energy
condition (NEC). The paper is planned as follows. In section
\textbf{2}, we employ the Karmarkar condition to formulate a WH
shape function. The field equations are constructed in section
\textbf{3} to give a brief overview of $f(R,T)$ theory. We take
three different viable models of this theory to examine the behavior
of NEC graphically. In section \textbf{4}, we study the stability of
a viable WH through the speed of sound and adiabatic index. We
summarize our results in section \textbf{5}.

\section{Wormhole and Karmarkar Condition}

In this section, we formulate the WH shape function through
Karmarkar condition which determines the structure of WH. For this
purpose, we consider a static spherically symmetric spacetime as
\begin{equation}\label{1}
ds^{2}=-e^{\mu(r)}dt^{2}+e^{\nu(r)}dr^{2}+r^{2}(d\theta^{2}
+\sin^{2}\theta d\phi^{2}).
\end{equation}
The corresponding non-zero Riemann curvature components are
\begin{eqnarray}\nonumber
R_{1212}&=&\frac{e^{\mu}(2\mu''+\mu'^{2}-\mu'\nu')}{4}, \quad
R_{3434}=\frac{r^{2}\sin^{2}\theta(e^{\nu}-1)}{e^{\nu}},
\\\nonumber
R_{1414}&=&\frac{r\sin^{2}\theta\mu'e^{\mu-\nu}}{2}, \quad
R_{2323}=\frac{r\nu'}{2}, \quad R_{1334}=R_{1224}\sin^{2}\theta.
\end{eqnarray}
The well-known Karmarkar constraint is fulfilled by the above
Riemann elements as
\begin{eqnarray}\label{2}
R_{1414}&=&\frac{R_{1212}R_{3434}+R_{1224}R_{1334}}{R_{2323}},\quad
R_{2323}\neq0.
\end{eqnarray}
The spacetime which satisfies the Karmarkar constraint is named as
embedding class-I. Substituting the values of Riemann components in
Eq.(\ref{2}), we obtain
\begin{equation}\nonumber
\frac{\mu'\nu'}{1-e^{\nu}}=\mu'\nu'-2\mu''-\mu'^{2},
\end{equation}
where $e^{\nu}\neq1$. The corresponding solution is
\begin{equation}\label{3}
e^{\nu}=1+A e^{\mu}\mu'^{2},
\end{equation}
where $A$ is an integration constant.

We assume the Morris-Thorne metric to describe the shape function as
\begin{equation}\label{4}
ds^{2}=-e^{\mu(r)}dt^{2}+\frac{1}{1-\frac{\epsilon(r)}{r}}dr^{2}
+r^{2}d\theta^{2}+r^{2}\sin\theta d\phi^{2},
\end{equation}
here $\mu(r)=\frac{-2\xi}{r}$ ($\xi$ is an arbitrary constant) is
considered the redshift function as $\mu(r)\rightarrow0$, when
$r\rightarrow\infty$ \cite{51}. By comparing Eqs.(\ref{1}) and
(\ref{4}), we have
\begin{equation}\label{5}
\nu(r)=\ln\bigg[\frac{r}{r-\epsilon(r)}\bigg].
\end{equation}
We evaluate the WH shape function from Eqs.(\ref{3}) and (\ref{5})
as
\begin{equation}\label{6}
\epsilon(r)=r-\frac{r^{5}}{r^{4}+4\xi^{2}Ae^{\frac{-2\xi}{r}}}.
\end{equation}
According to Morris and Thorne \cite{5}, the shape function must
fulfill the following conditions to obtain a traversable WH
solution.
\begin{enumerate}
\item
$\epsilon(r)<r$,
\item
$\epsilon(r)-r=0$ at $r=a$,
\item
$\frac{\epsilon(r)-r\epsilon'(r)}{\epsilon^{2}(r)}>0$ at $r=a$,
\item
The condition $\epsilon'(r)<1$ should be satisfied,
\item
$\frac{\epsilon(r)}{r}\rightarrow0$ must satisfy as
$r\rightarrow\infty$,
\end{enumerate}
where $a$ is the WH throat radius. Equation (\ref{6}) has a trivial
solution at WH throat, i.e., $\epsilon(a)-a=0$. To obtain a
non-trivial solution, we modify Eq.(\ref{6}) with the inclusion of
free parameter $c$ as
\begin{eqnarray}\label{6a}
\epsilon(r)=r-\frac{r^{5}}{r^{4}+4\xi^{2}Ae^{\frac{-2\xi}{r}}}+c.
\end{eqnarray}
The value of $c$ should be confined in the range $0<c<a$. For other
values of $c$, the conditions of wormhole shape function are not
satisfied that are necessary for a physically viable WH structures.
Using the condition $\epsilon(r)-r=0$ at $r=a$ in the above
equation, we obtain
$A=\frac{a^{4}(a-c)}{4\xi^{2}e^{\frac{-2\xi^{2}}{a}}}$. Substituting
the value of $A$ in the above equation, the shape function becomes
\begin{eqnarray}\label{7}
\epsilon (r)=r-\frac{r^{5}}{r^{4}+a^{4}(a-c)}+c.
\end{eqnarray}
This can easily be seen that this shape function satisfies the
condition (5) as follows
\begin{equation}\label{8}
\lim _{r\rightarrow\infty}\frac{\epsilon(r)}{r}=0.
\end{equation}
Thus, asymptotically flat traversable WHs are obtained through this
WH shape function. For our convenience, we consider $a=2$ and
$\xi=-1$ in all the graphs. The graphical behavior of the WH shape
function is given in Figure \textbf{1} which shows that the WH shape
function satisfies all the necessary conditions.
\begin{figure}
\epsfig{file=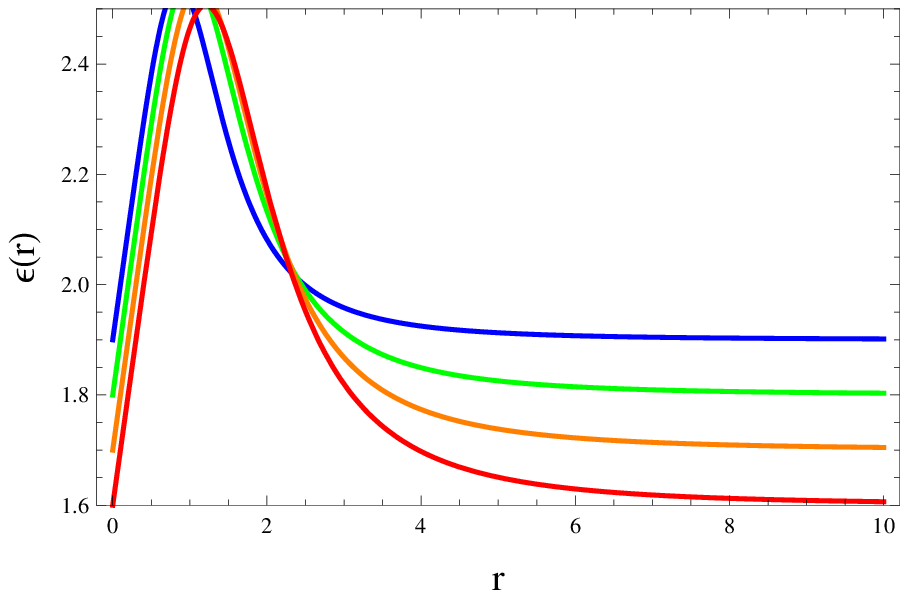,width=.5\linewidth}
\epsfig{file=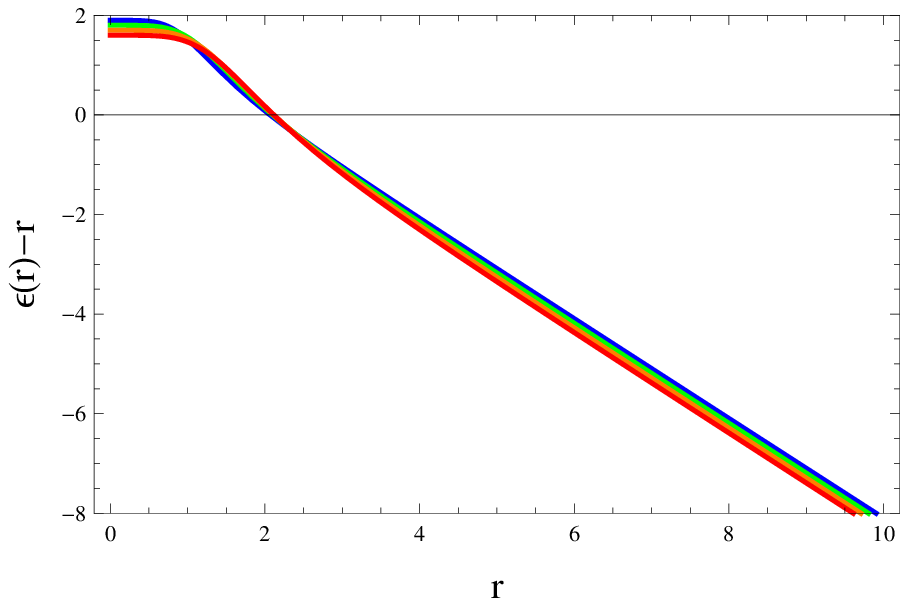,width=.5\linewidth}
\epsfig{file=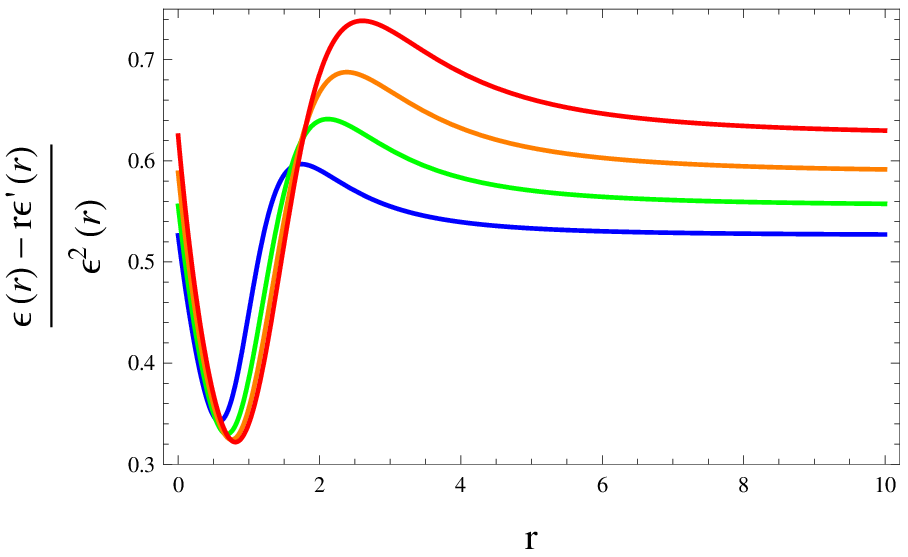,width=.5\linewidth}
\epsfig{file=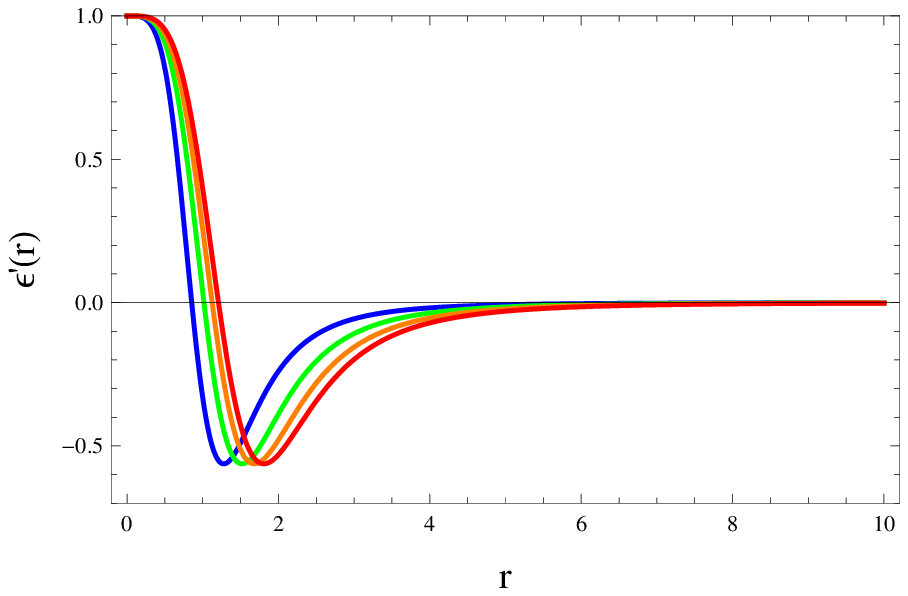,width=.5\linewidth}\center
\epsfig{file=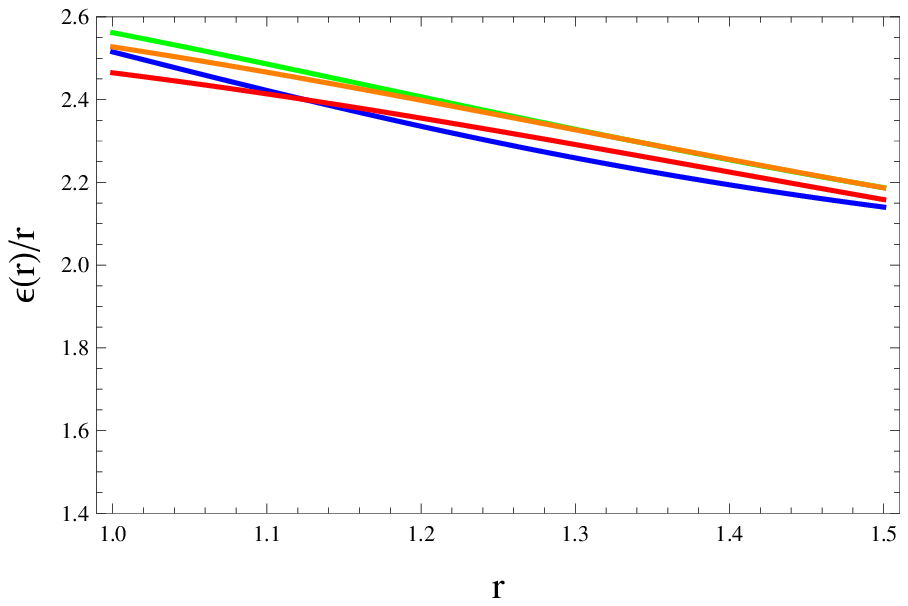,width=.5\linewidth}\caption{Behavior of shape
function corresponding to radial coordinate for $c=1.9$ (blue line),
$c=1.8$ (green line), $c=1.7$ (orange line) and $c=1.6$ (red line).}
\end{figure}

\section{Basic Formalism of $f(R,T)$ Gravity}

The action of this modified theory is defined as
\begin{equation}\label{9}
I=\int dx^{4}\sqrt{-g}[f(R,T)+L_{m}],
\end{equation}
where $g$ represents determinant of the line element and $L_{m}$ is
the matter-Lagrangian density. The corresponding field equations are
\begin{equation}\label{10}
R_{\alpha\beta}f_{R}-\frac{1}{2}g_{\alpha\beta}f
+(g_{\alpha\beta}\Box-\nabla_\alpha\nabla_\beta)f_{_{R}}+f_{T}\Theta
_{\alpha\beta}+f_{T}T_{\alpha\beta}=T_{\alpha\beta}.
\end{equation}
Here, $f\equiv f(R,T)$ and $f_{R}$ represents derivative with
respect to the Ricci scalar and $f_{T}$ denotes derivative with
respect to trace of the energy-momentum tensor. The expression of
$\Theta_{\alpha\beta}$ is given by
\begin{equation}\label{10a}
\Theta
_{\alpha\beta}=-2T_{\alpha\beta}+g_{\alpha\beta}L_{m}-2g^{\upsilon\tau}
\frac{\partial^{2}L_{m}}{\partial g^{\alpha\beta}\partial
g^{\upsilon\tau}}.
\end{equation}
The $f(R,T)$ gravity provides non-conserved stress-energy tensor
implying the presence of an extra force which acts as a non-geodesic
motion of particles given by
\begin{equation}\nonumber
\nabla^{\alpha}T_{\alpha\beta}=\frac{f_{T}}{1-f_{T}}\bigg[(T_{\alpha\beta}
+\Theta_{\alpha\beta})\nabla^{\alpha}\ln
f_{T}+\nabla^{\alpha}\Theta_{\alpha\beta}-\frac{1}{2}g_{\alpha\beta}T\bigg].
\end{equation}
This shows that the covariant derivative of the stress-energy tensor
in $f(R,T)$ theory does not vanish.

We consider anisotropic fluid configuration as
\begin{equation}\label{11}
T_{\alpha\beta}=(P_{r}-P_{t})\chi_{\alpha}\chi_{\beta}-P_{t}g_{\alpha
\beta}+(\rho+P_{t})V_{\alpha}V_{\beta},
\end{equation}
where $\rho$, $P_{r}$ and $P_{t}$ represent the energy density,
radial and tangential pressures, respectively. Using Eqs.(\ref{1})
and (\ref{10}), the resulting field equations become
\begin{eqnarray}\label{14}
\rho&=&\frac{1}{e^{\nu}}\bigg[-\frac{f}{2}e^{\nu}+\big(\frac{\mu'}{r}
-\frac{\mu'\nu'}{4}+\frac{\mu''}{2}+\frac{\mu'^{2}}{4}\big)f_{R}
+\big(\frac{\nu'}{2}-\frac{2}{r}\big)f'_{R}-f''_{R}\bigg],
\\\nonumber
P_{r}&=&\frac{1}{{e^{\nu}(1+f_{T})}}\bigg[\frac{f}{2}e^{\nu}+\big
(\frac{\mu''}{2}-\frac{\nu'}{r}-\frac{\mu'\nu'}{4}+\frac{\mu'^{2}}
{4}\big)f_{R}+\big(\frac{\mu'}{2}+\frac{2}{r}\big)f'_{R}
\\\label{15}
&-&\rho f_{T}e^{\nu}\bigg],
\\\nonumber
P_{t}&=&\frac{1}{{e^{\nu}(1+f_{T})}}\bigg[\frac{f}{2}e^{\nu}+\big
(\frac{(\nu'-\mu')r}{2}+e^{\nu}-1\big)\frac{f_{R}}{r^{2}}+\big
(\frac{\mu'-\nu'}{2}+\frac{1}{r}\big) f'_{R}
\\\label{16}
&+&f''_{R}-\rho f_{T}e^{\nu}\bigg].
\end{eqnarray}

The field equations (\ref{14})-(\ref{16}) appear to be more complex
due to the existence of multivariate functions and their
derivatives. We assume a particular $f(R,T)$ model to solve these
equations as \cite{23}
\begin{equation}\label{17}
f(R,T)=f_{1}(R)+f_{2}(T).
\end{equation}
Different viable models of $f(R,T)$ theory can be discussed by
considering various forms of $f_{1}(R)$ and $f_{2}(T)$. However, for
the sake of simplicity, we take $f_{2}(T)=\lambda T$ \cite{23} to
solve the field equations. There can be other choices of $f(T)$ such
that one can take non-minimal model. The field equations
corresponding to the model (\ref{17}) are given in
Eqs.(\ref{A1})-(\ref{A3}) of Appendix \textbf{A}.

Now, we study various $f(R,T)$ models corresponding to $f_{1}(R)$ in
the following subsections.

\subsection{Model 1: Exponential Gravity Model}

Cognola et al \cite{52} investigated the exponential gravity model
as
\begin{equation}\label{21}
f(R)=R-KB\bigg[1-e^{\frac{-R}{B}}\bigg],
\end{equation}
which explains the current cosmic expansion as well as inflation of
the early time. Here $K$ and $B$ are arbitrary constants. The
resulting equations of motion are given in Eqs.(\ref{A4})-(\ref{A6})
of Appendix \textbf{A}. Energy constraints play a significant role
to determine the physical existence of cosmic structures. These
energy bounds must be violated for the existence of realistic WH
geometry. In modified theories, the violation of NEC $(\rho+P_{t}$,
$\rho+P_{r})$ ensures the existence of a viable traversable WH
geometry. The graphical behavior of NEC for different values of $K$
and $B$ is given in Figure \textbf{2}. This shows that matter
variables violate the null energy bound for positive values of $B$
and negative values of $K$ which ensures the presence of viable
traversable WH. However, positive values of model parameters satisfy
the NEC and hence show the absence of exotic matter at WH throat.
\begin{figure}
\epsfig{file=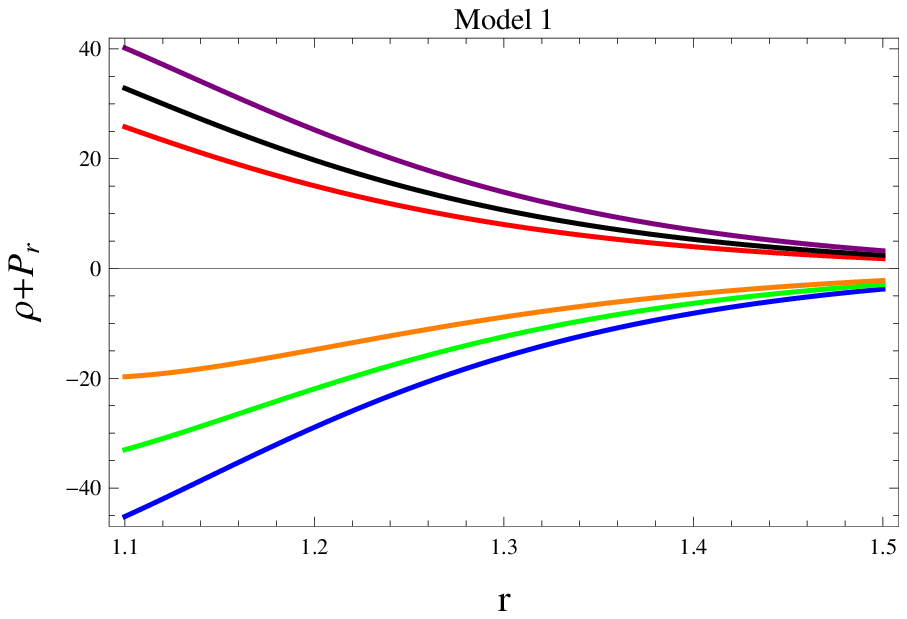,width=.5\linewidth}
\epsfig{file=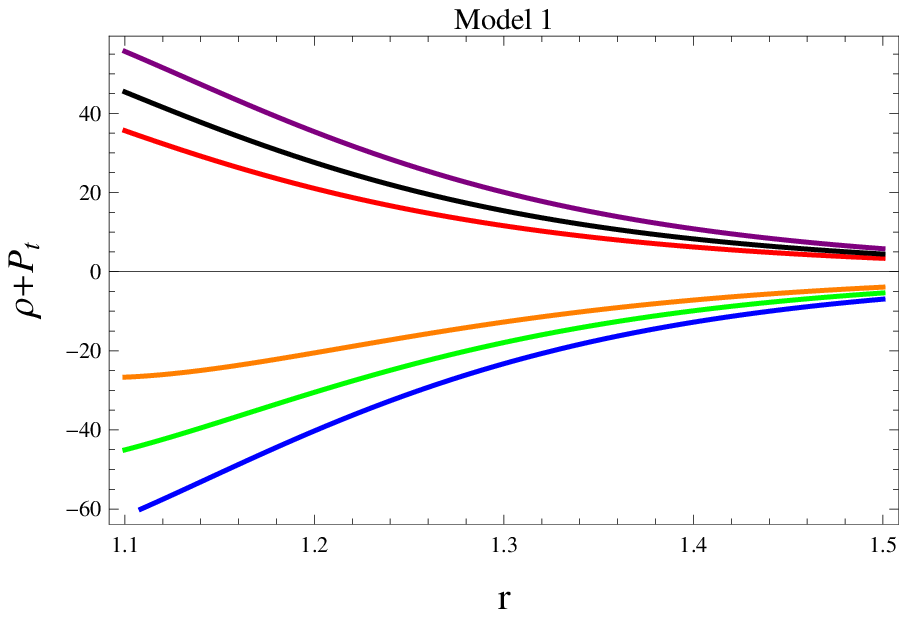,width=.5\linewidth}\caption{NEC versus radial
coordinate for $K=-7$, $B=9$ (blue line), $K=-5$, $B=8$ (green
line), $K=-3$, $B=6$ (orange line), $K=5$, $B=14$ (red line), $K=6$,
$B=12$ (black line), $K=7$, $B=10$ (purple line).}
\end{figure}

\subsection{Model 2: Starobinsky Gravity Model}

Here, we use the Starobinsky model \cite{53} as
\begin{equation}\label{22}
f(R)=R-\gamma B\bigg[1-\bigg(1+\frac{R^{2}}{B^{2}}\bigg)^{-n}\bigg],
\end{equation}
which satisfies the solar system tests, complies all the cosmic as
well as localized gravity restrictions and has attributes of dark
energy models, $\gamma$ and $n$ are constants. The field equations
corresponding to this model are given in Eqs.(\ref{A7})-(\ref{A9})
of Appendix \textbf{A}. The graphical behavior of NEC for positive
as well as negative values of $n$ is given in Figure \textbf{3}. For
$n<0$, the NEC violates corresponding to negative values of $B$ and
positive values of $\gamma$ which gives the presence of traversable
spherically symmetric WH geometry. For $n>0$, when $B<0$ and
$\gamma<0$, a small amount of exotic matter is present at the WH
throat so a viable WH can be obtained.
\begin{figure}
\epsfig{file=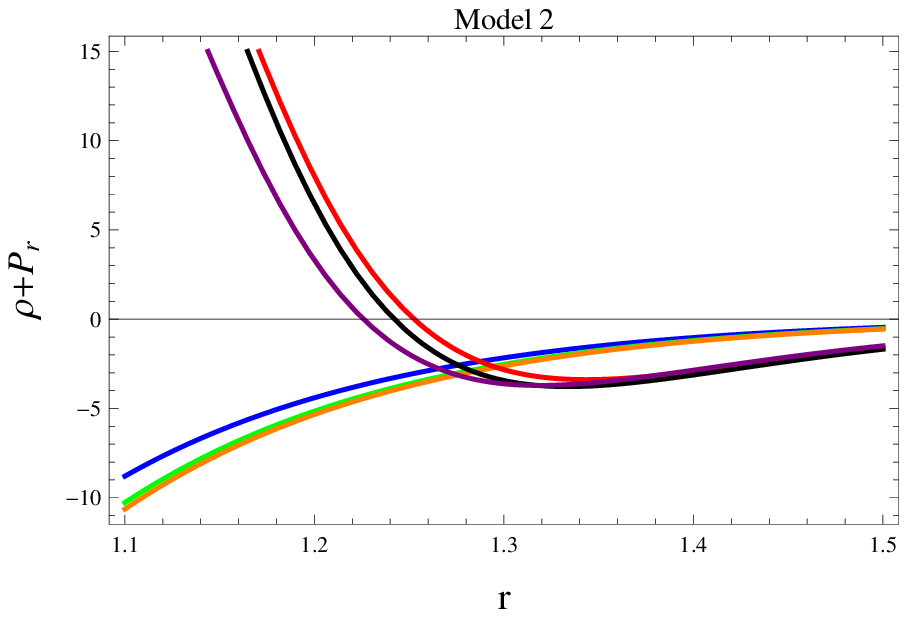,width=.5\linewidth}
\epsfig{file=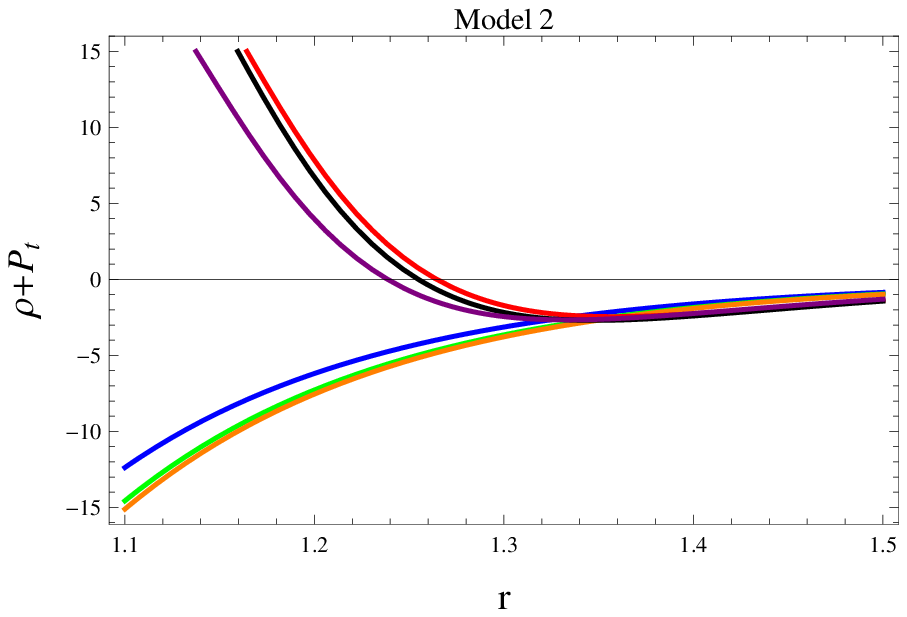,width=.5\linewidth}\caption{NEC versus radial
coordinate for $\gamma=1$, $B=-23$, $n=-1$ (blue line),
$\gamma=1.5$, $B=-25$, $n=-0.9$ (green line), $\gamma=1.8$, $B=-27$,
$n=-0.85$ (orange line), $\gamma=-0.6$, $B=-5.4$, $n=2$ (red line),
$\gamma=-0.8$, $B=-5.2$, $n=1.5$ (black line), $\gamma=-1$, $B=-5$,
$n=1$ (purple line).}
\end{figure}

\subsection{Model 3: Tsujikawa Gravity Model}

Finally, we consider another well-known model \cite{54}, defined as
\begin{equation}\label{23}
f(R)=R-\omega B\tanh\bigg(\frac{R}{B}\bigg),
\end{equation}
where $\omega$ is an arbitrary constant and the corresponding field
equations are given in Eqs.(\ref{A10})-(\ref{A12}) of Appendix
\textbf{A}. Figure \textbf{4} shows the graphical behavior of NEC
for different values of $\omega$ and $B$. This indicates the
existence of viable traversable WH for negative values of $B$ and
$\omega$. However, positive values of $\omega$ and negative values
of $B$ satisfy the NEC.
\begin{figure}
\epsfig{file=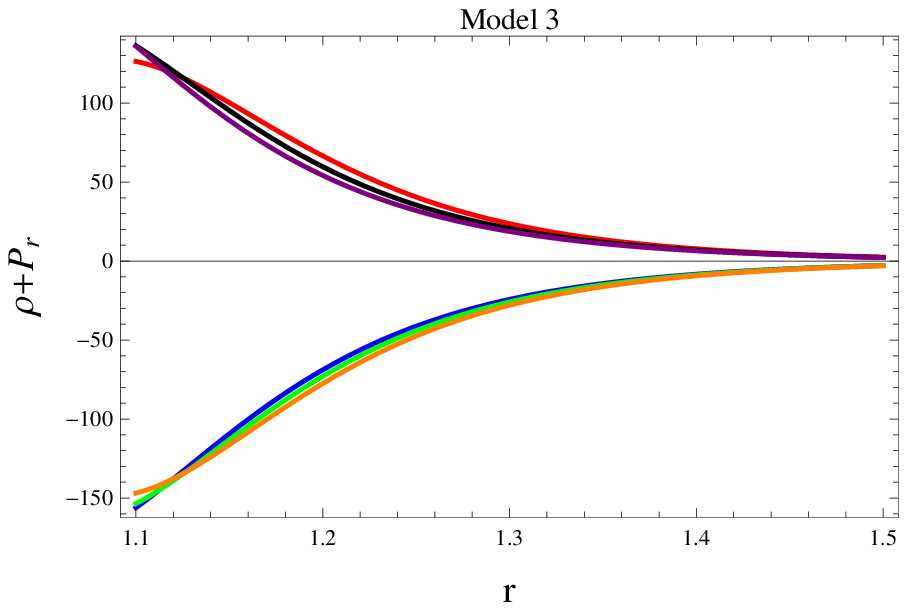,width=.5\linewidth}
\epsfig{file=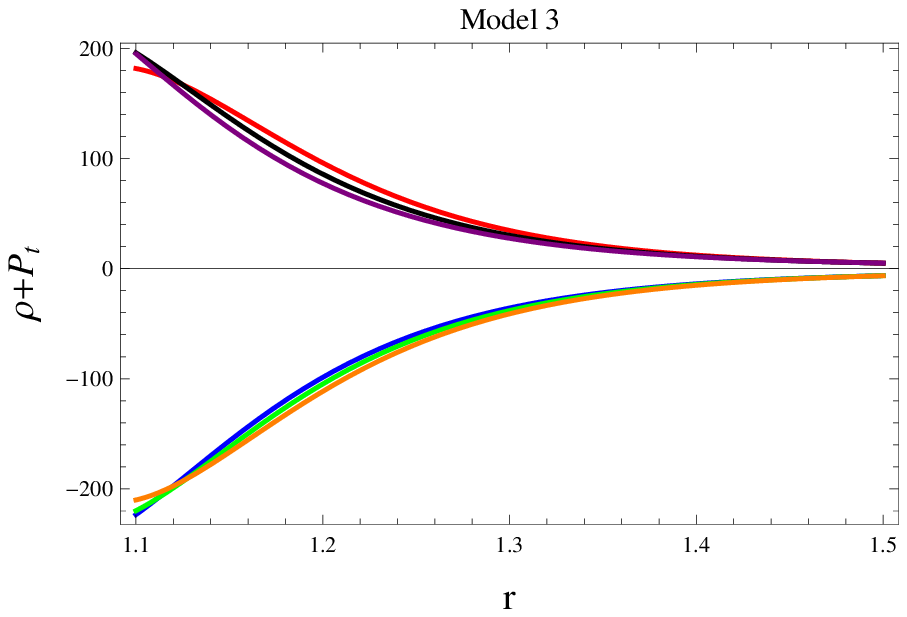,width=.5\linewidth}\caption{NEC versus $r$ for
$\omega=-10$, $B=-6$, (blue line), $\omega=-9.5$, $B=-5.5$, (green
line), $\omega=-9$, $B=-5$, (orange line), $\omega=8$, $B=-5.2$,
(red line), $\omega=9$, $B=-6$, (black line), $\omega=10$, $B=-7$,
(purple line).}
\end{figure}

\section{Stability Analysis}

Stability is significant to analyze the valid and consistent cosmic
structures. It is more interesting to examine cosmic objects that
display stable behavior under the external perturbations. Here, we
examine the stability of viable and traversable WH solutions through
sound speed and adiabatic index.

\subsection{Speed of Sound}

We investigate the stability of WH solution through causality
condition and Herrera's cracking method. The causality criteria
states that the speed of sound components
$(v^{2}_{r}=\frac{dP_{r}}{d\rho},~v_{t}^{2} =\frac{dP_{t}}{d\rho})$
should be confined in the range [0,1] for stable structures. Figure
\textbf{5} shows that traversable WH satisfy the required causality
condition in the presence of modified terms. According to Herrera
cracking technique, the difference in sound speed components should
be $0\leq\mid v_{t}^{2}-v^{2}_{r}\mid\leq1$.  Figure \textbf{6}
demonstrates that WH solutions are in the stable state as they
satisfy the required condition corresponding to all considered
$f(R,T)$ models. Hence, we obtain physically stable traversable WH
in this modified theory.
\begin{figure}
\epsfig{file=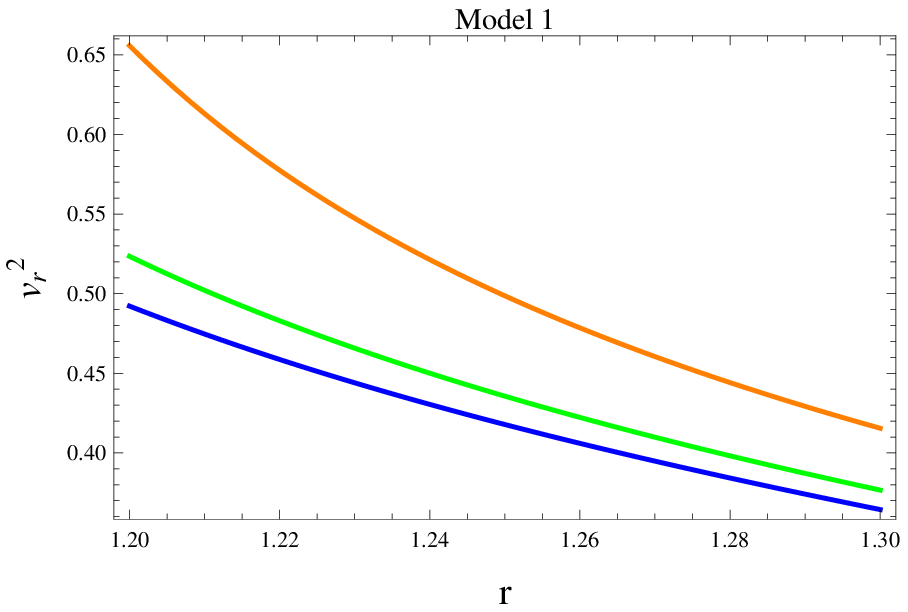,width=.5\linewidth}
\epsfig{file=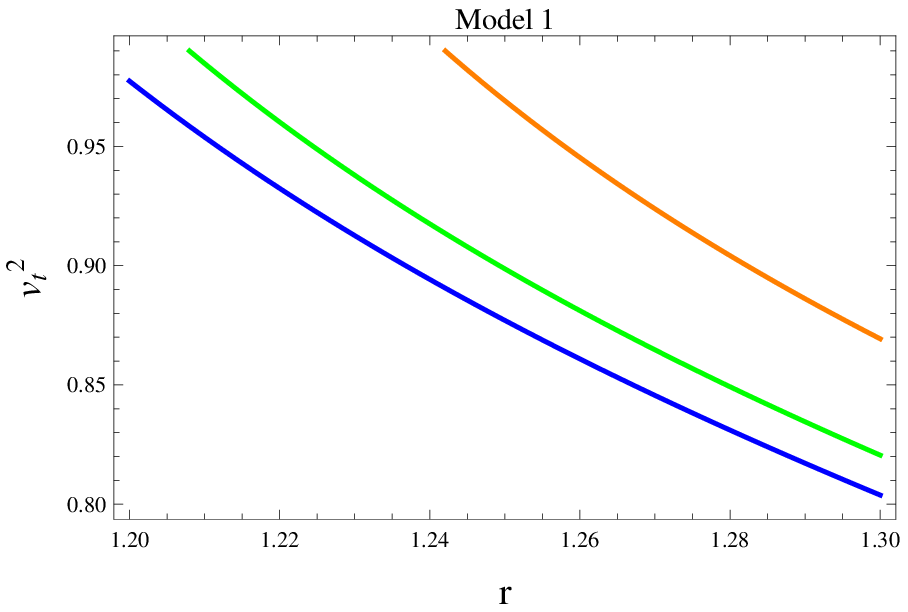,width=.5\linewidth}
\epsfig{file=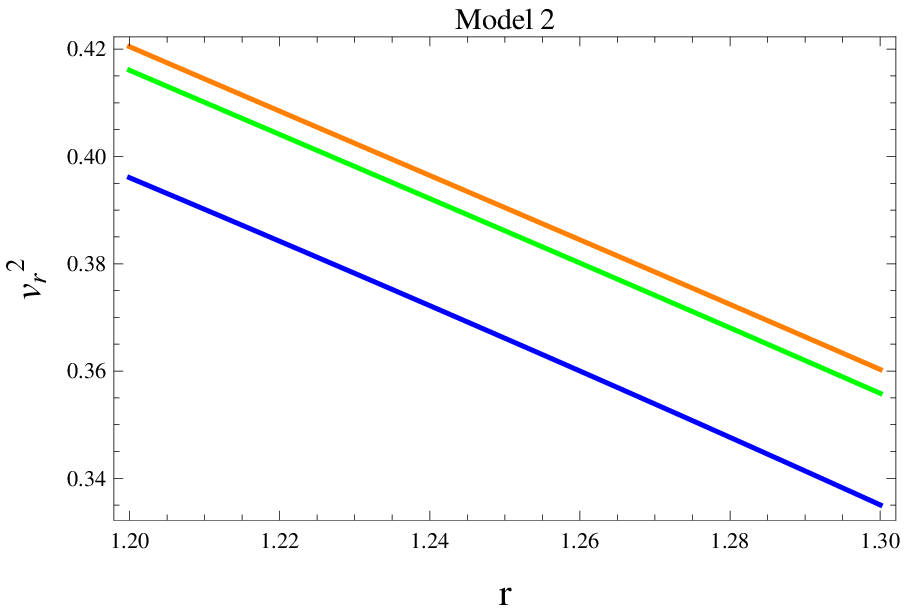,width=.5\linewidth}
\epsfig{file=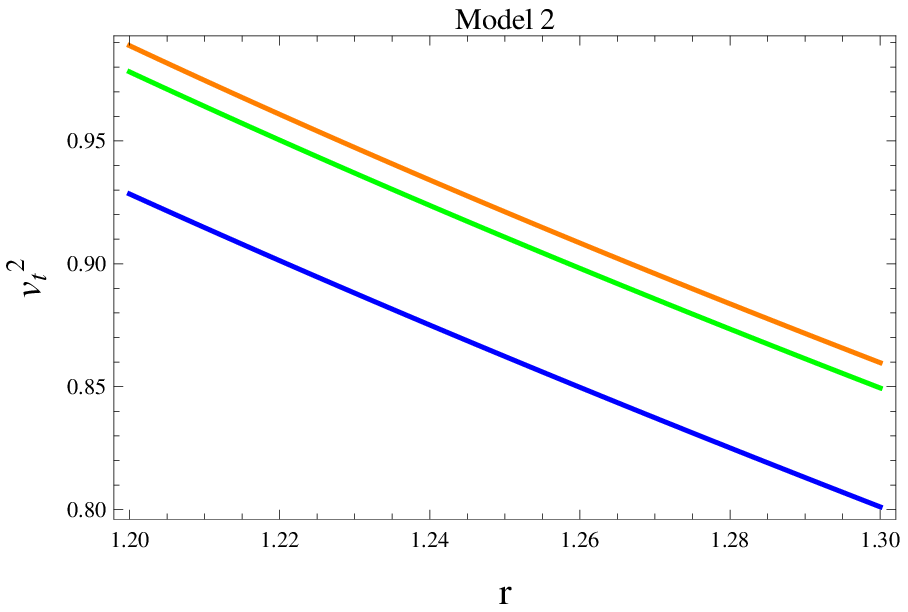,width=.5\linewidth}
\epsfig{file=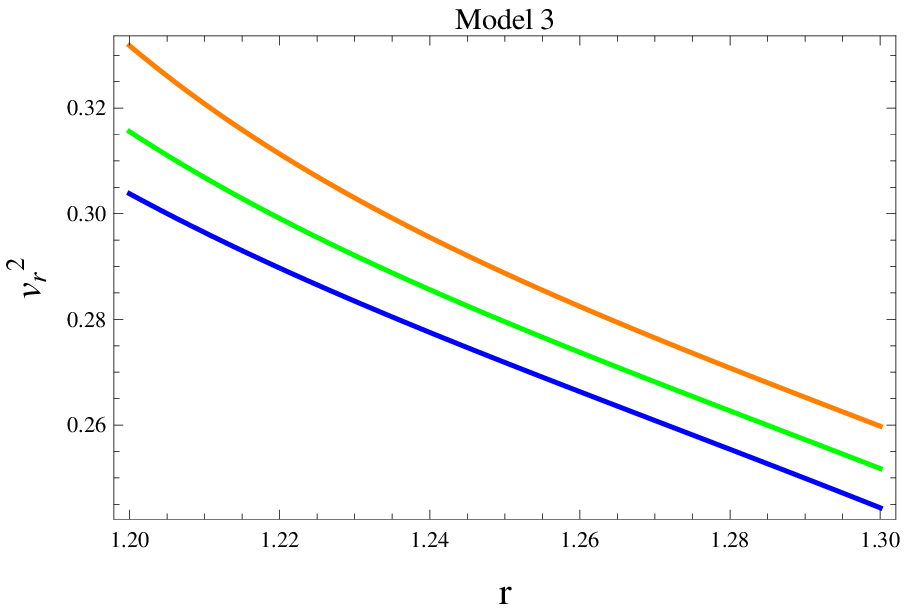,width=.5\linewidth}
\epsfig{file=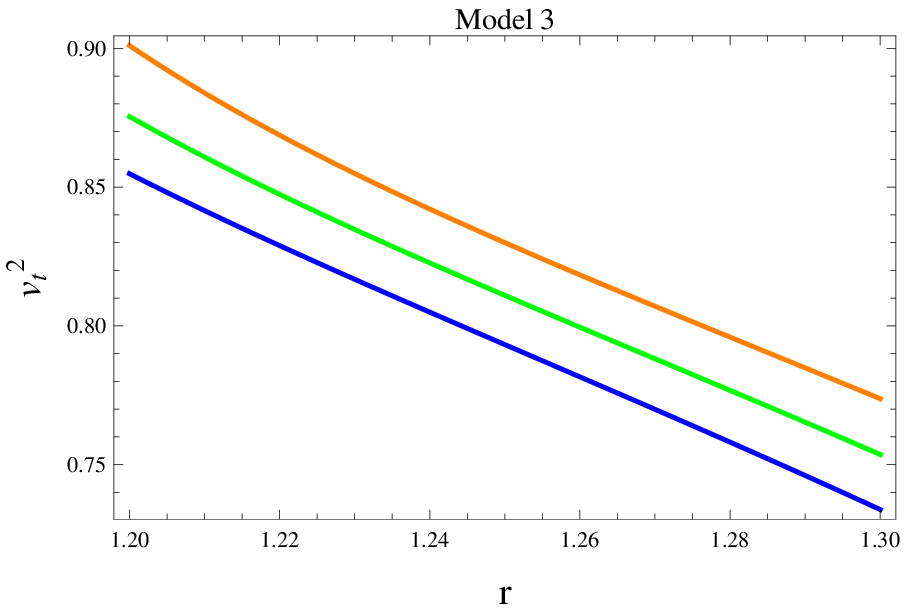,width=.5\linewidth}\caption{Behavior of
causality condition versus $r$.}
\end{figure}
\begin{figure}
\epsfig{file=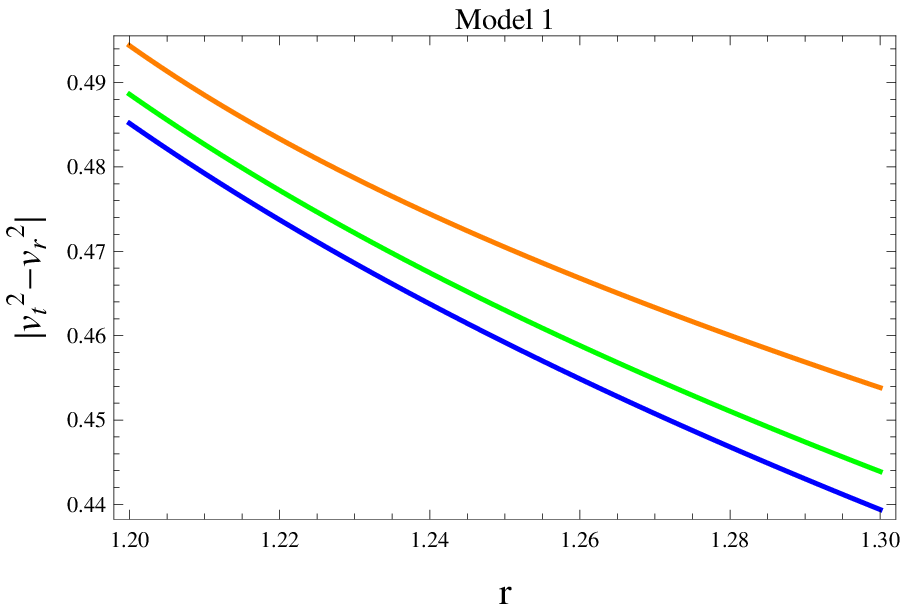,width=.5\linewidth}
\epsfig{file=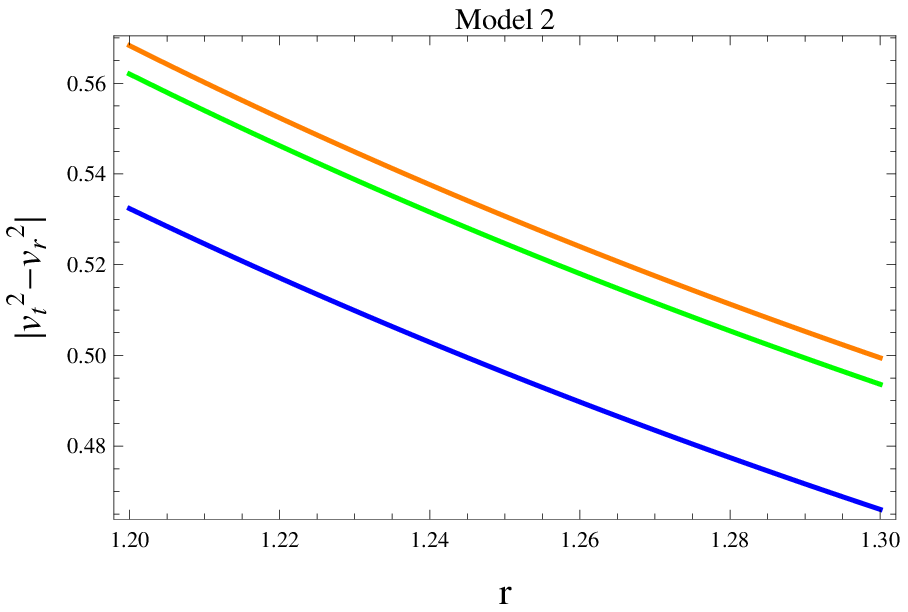,width=.5\linewidth} \center
\epsfig{file=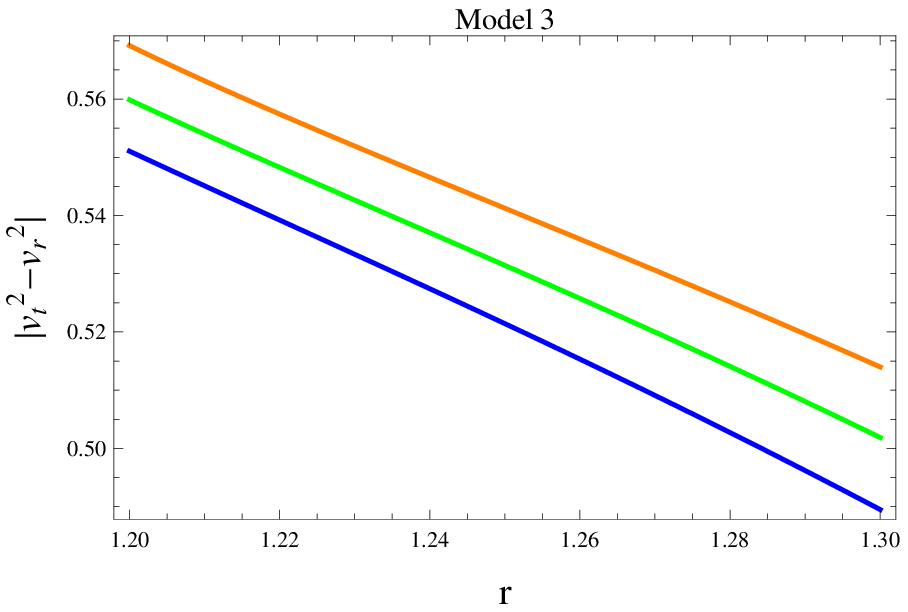,width=.5\linewidth}\caption{Behavior of
Herrera cracking versus $r$.}
\end{figure}

\subsection{Adiabatic Index}
\begin{figure}
\epsfig{file=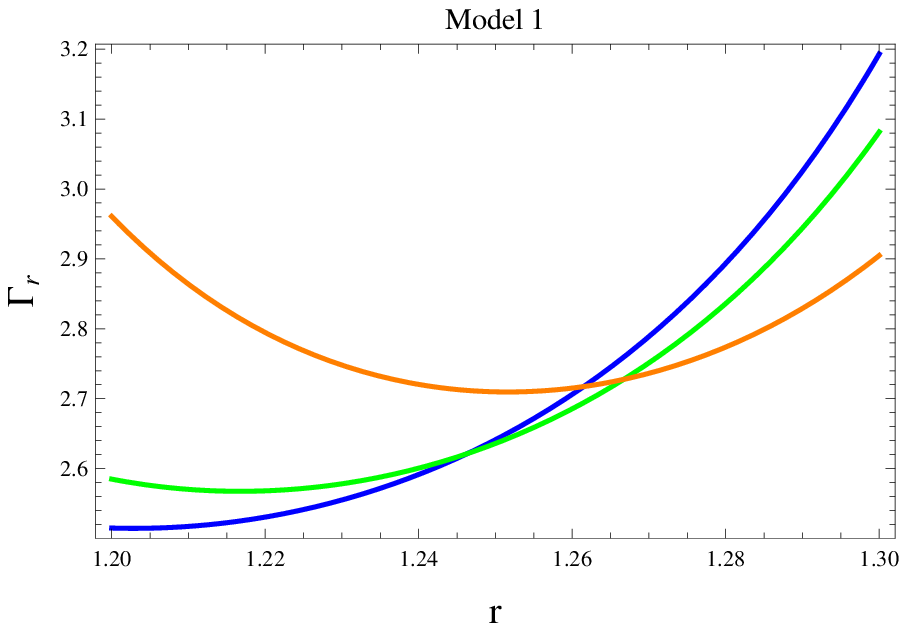,width=.5\linewidth}
\epsfig{file=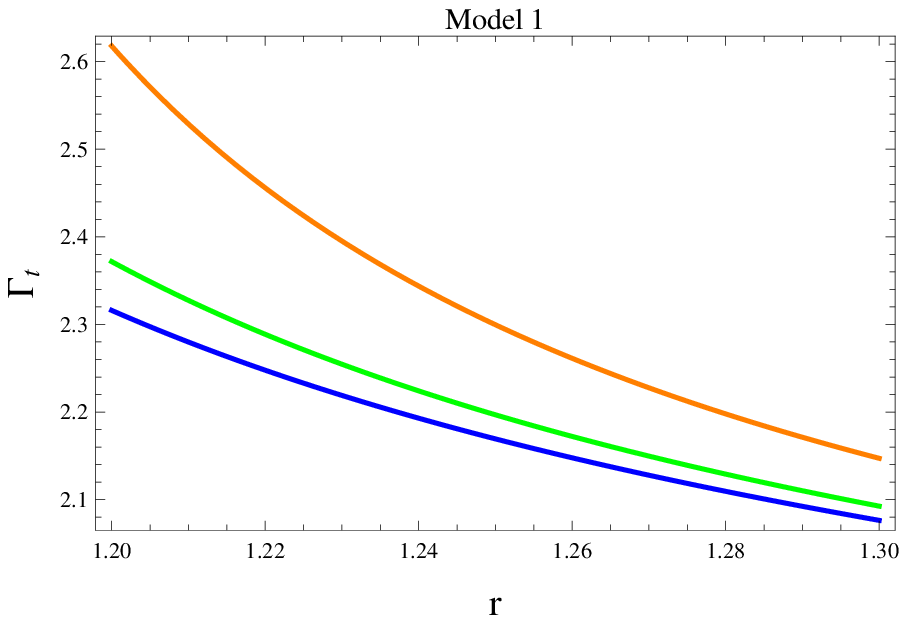,width=.5\linewidth}
\epsfig{file=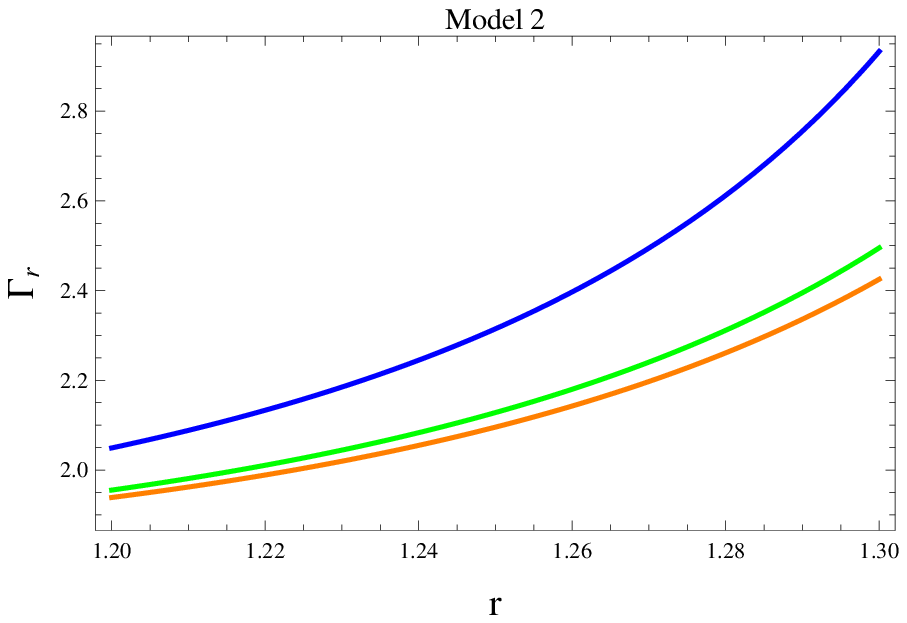,width=.5\linewidth}
\epsfig{file=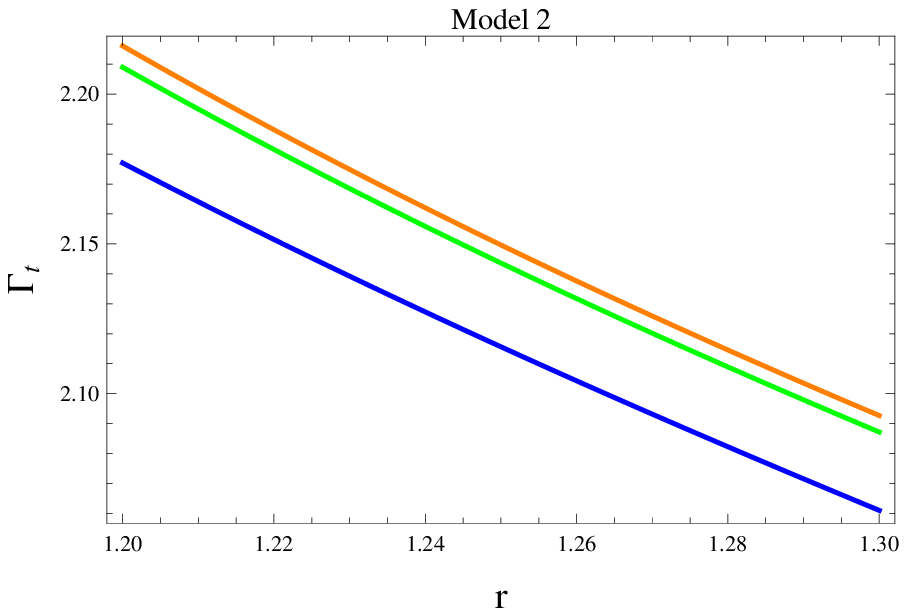,width=.5\linewidth}
\epsfig{file=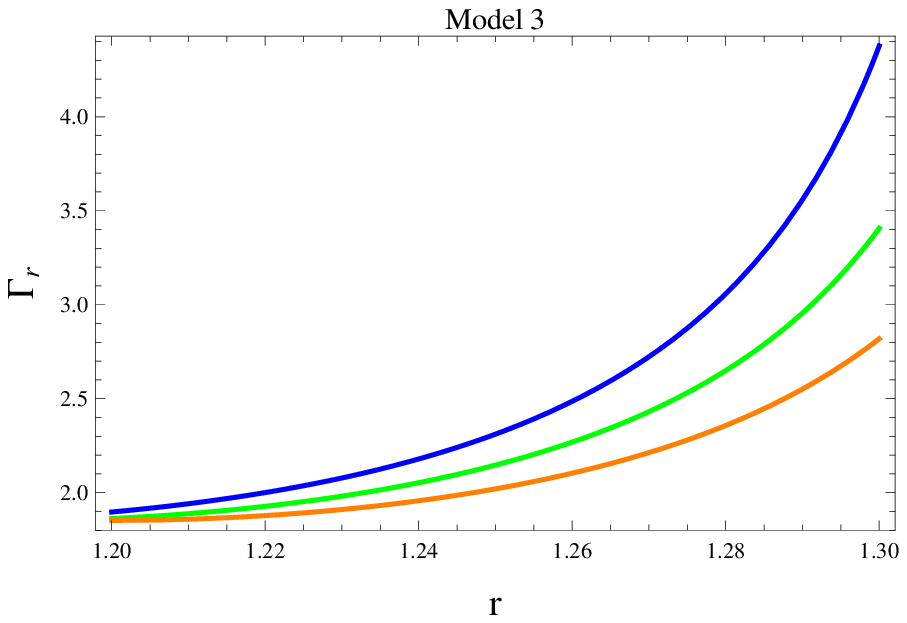,width=.5\linewidth}
\epsfig{file=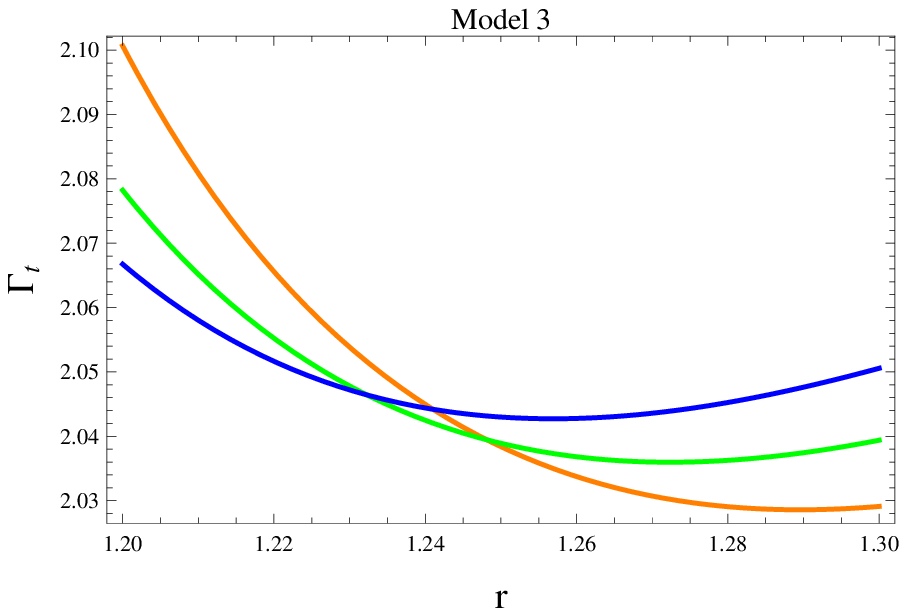,width=.5\linewidth}\caption{Behavior of
adiabatic index versus $r$.}
\end{figure}

The adiabatic index is another factor used to explore the stability
of celestial objects. The radial and transverse components of
adiabatic index for anisotropic fluid are defined as
\begin{eqnarray}\nonumber
\Gamma_{r}=\frac{\rho+P_{r}}{P_{r}}\frac{dP_{r}}{d\rho},\quad
\Gamma_{t}=\frac{\rho+P_{t}}{P_{t}}\frac{dP_{t}}{d\rho}.
\end{eqnarray}
According to Heintzmann and Hillebrandt \cite{55}, a system is
stable if $\Gamma>4/3$, otherwise it is unstable. Figure \textbf{7}
shows that traversable WH solutions have adiabatic index greater
than $4/3$, indicating that our system is stable even when
higher-order matter source terms are present.

\section{Final Remarks}

There have been different methods to obtain viable WH solutions in
the literature. One of them is to evaluate the shape function by
making certain hypotheses for the matter ingredients and the other
is to investigate how the energy conditions behave by considering
the shape function. In this paper, we have examined whether WH
solutions exist or not by building the shape function using Karmakar
condition in $f(R,T)$ theory. We have explored exact solutions of
static spherically symmetric traversable WHs corresponding to three
viable $f(R,T)$ models, i.e., the exponential gravity model,
Starobinsky gravity model and Tsujikawa gravity model. We have
checked the graphical behavior of NEC to examine the traversable WH
geometry. The stability is examined through the causality condition,
Herrera cracking approach and the adiabatic index. The summary of
the obtained results is given as follows.
\begin{itemize}
\item
The considered shape function yields a viable WH structure by
satisfying all the required conditions as shown in Figure
\textbf{1}.
\item
The exponential gravity model shows that NEC violates for $K<0$ and
$B>0$, which ensures the presence of exotic matter at WH throat.
Hence, viable traversable WH is obtained for specific values of the
model parameters.
\item
For the Starobinsky gravity model, WH solutions exist for a wide
range of parameters, i.e., for $B<0$, $n<0$ and $\gamma>0$ which
shows the existence of exotic matter at WH throat (Figure
\textbf{3}). When $n>0$ and $\gamma<0$, NEC is again violated with
negligible amount of exotic matter. Thus we have found viable
traversable WH with suitable values of the model parameters.
\item
The presence of exotic matter at the throat is confirmed for the
Tsujikawa gravity model when $\omega<0$ and $B<0$.
\item
The stability requirements are verified in the presence of modified
terms (Figures \textbf{5-7}).
\end{itemize}

In the framework of GR, Fayyaz and Shamir \cite{22} found viable and
stable traversable WH structure in the presence of exotic matter.
The same authors \cite{48} concluded that there exist WH solution
with a negligible amount of exotic matter in $f(R)$ theory for the
specific shape function. It is worth mentioning here that we have
obtained WH solutions in $f(R,T)$ gravity as NEC is violated which
ensures the presence of exotic matter at WH throat. Thus we can
conclude that viable and stable traversable WH solutions exist for
anisotropic matter configuration through Karmarkar condition in this
modified theory.

\vspace{0.25cm}

\section*{Appendix A}
\renewcommand{\theequation}{A\arabic{equation}}
\setcounter{equation}{0}

The field equations corresponding to the model (\ref{17}) are
\begin{eqnarray}\nonumber
\rho&=&\frac{1}{{e^{\nu}}2(2\lambda+1)(\lambda+1)}\bigg[(5\lambda+2)
\bigg\{-\frac{f}{2}e^{\nu}+\big(\frac{\mu'}{r}-\frac{\mu'\nu'}{4}
+\frac{\mu''}{2}
\\\nonumber
&+&\frac{\mu'^{2}}{4}\big)f_{R}+\big(\frac{\nu'}{2}-\frac{2}{r}\big)
f'_{R}-f''_{R}\bigg\}+\lambda\bigg\{\frac{f}{2}e^{\nu}+\big(\frac
{\mu''}{2}-\frac{\nu'}{r}-\frac{\mu'\nu'}{4}
\\\nonumber
&+&\frac{\mu'^{2}}{4}\big)f_{R}
+\big(\frac{\mu'}{2}+\frac{2}{r}\big)f'_{R}\bigg\}+2\lambda
\bigg\{\frac{f}{2}e^{\nu}-\big(\frac{(\mu'-\nu')r}{2}-e^{\nu}+1\big)
\\\label{A1}
&\times&\frac{f_{R}}{r^{2}}+\big(\frac{\mu'-\nu'}{2}+\frac{1}{r}\big)
f'_{R}+f''_{R}\bigg\}\bigg],
\\\nonumber
P_{r}&=&\frac{1}{{e^{\nu}}2(2\lambda+1)(\lambda+1)}\bigg[-\lambda\bigg
\{-\frac{f}{2}e^{\nu}+\big(\frac{\mu'}{r}-\frac{\mu'\nu'}{4}
+\frac{\mu''}{2}
\\\nonumber
&+&\frac{\mu'^{2}}{4}\big)f_{R}+\big(\frac{\nu'}{2}-\frac{2}{r}\big)
f'_{R}-f''_{R}\bigg\}+(3\lambda+2)\bigg\{\frac{f}{2}e^{\nu}+\big(\frac
{\mu''}{2}-\frac{\nu'}{r}
\\\nonumber
&-&\frac{\mu'\nu'} {4}+\frac{\mu'^{2}}{4}\big)f_{R}
+\big(\frac{\mu'}{2}+\frac{2}{r}\big)f'_{R}\bigg\}-2\lambda
\bigg\{\frac{f}{2}e^{\nu}+\big(\frac{(\mu'-\nu')r}{2}
\\\label{A2}
&-&e^{\nu}+1\big)
\frac{-f_{R}}{r^{2}}+\big(\frac{\mu'-\nu'}{2}+\frac{1}{r}\big)f'_{R}
+f''_{R}\bigg\}\bigg],
\\\nonumber
P_{t}&=&\frac{1}{{e^{\nu}}2(2\lambda+1)(\lambda+1)}\bigg[-\lambda\bigg
\{-\frac{f}{2}e^{\nu}+\big(\frac{\mu'}{r}-\frac{\mu'\nu'}{4}+\frac
{\mu''}{2}
\\\nonumber
&+&\frac{\mu'^{2}}{4}\big)f_{R}
+\big(\frac{\nu'}{2}-\frac{2}{r}\big)f'_{R}-f''_{R}\bigg\}+\lambda
\bigg\{\frac{f}{2}e^{\nu}+\big(\frac{\mu''}{2}-\frac{\nu'}{r}-\frac
{\mu'\nu'}{4}
\\\nonumber
&+&\frac{\mu'^{2}}{4}\big)f_{R}
+\big(\frac{\mu'}{2}+\frac{2}{r}\big)f'_{R}\bigg\}-2(\lambda+1)
\bigg\{\frac{f}{2}e^{\nu}-\big(\frac{(\mu'-\nu')r}{2}-e^{\nu}+1\big)
\\\label{A3}
&\times&\frac{f_{R}}{r^{2}}+\big(\frac{\mu'-\nu'}{2}+\frac{1}{r}\big)
f'_{R}+f''_{R}\bigg\}\bigg].
\end{eqnarray}
The resulting field equations corresponding to the model 1 turn out
to be
\begin{eqnarray}\nonumber
\rho&=&\frac{1}{{e^{\nu}}2(2\lambda+1)(\lambda+1)}\bigg[(5\lambda+2)\bigg
\{-\frac{f}{2}e^{\nu}+\big(\frac{\mu'}{r}-\frac{\mu'\nu'}{4}+\frac
{\mu''}{2}+\frac{\mu'^{2}}{4}\big)
\\\nonumber
&\times&\big(1-K
e^{\frac{-R}{B}}\big)+\big(\frac{\nu'}{2}-\frac{2}{r}\big)\big(\frac{1}{B}K
e^{\frac{-R}{B}}\big)R' -\big\{\big(\frac{1}{B}K
e^{\frac{-R}{B}}\big)R''-\big(\frac{1}{B^{2}}K e^{\frac{-R}{B}}\big)
\\\nonumber
&\times&R'^{2}\big\}\bigg\}+\lambda
\bigg\{\frac{f}{2}e^{\nu}+\big(\frac{\mu''}{2}-\frac{\nu'}{r}-\frac
{\mu'\nu'}{4}+\frac{\mu'^{2}}{4}\big)(1-Ke^{\frac{-R}{B}})+\big
(\frac{\mu'}{2}+\frac{2}{r}\big)
\\\nonumber
&\times&\big(\frac{1}{B}Ke^{\frac{-R}{B}}\big)R'\bigg\}+2\lambda
\bigg\{\frac{f}{2}e^{\nu}+\big(\frac{(\mu'-\nu')r}{2}-e^{\nu}+1\big)
\frac{(-1+K e^{\frac{-R}{B}})}{r^{2}}
\\\nonumber
&+&\big(\frac{\mu'-\nu'}{2}+\frac{1}{r}\big)\big(\frac{1}{B}K
e^{\frac{-R}{B}}\big)R'+\big\{\big(\frac{1}{B}K
e^{\frac{-R}{B}}\big)R''
\\\label{A4}
&+&\big(\frac{-1}{B^{2}}K
e^{\frac{-R}{B}}\big)R'^{2}\big\}\bigg\}\bigg],
\\\nonumber
P_{r}&=&\frac{1}{{e^{\nu}}2(2\lambda+1)(\lambda+1)}\bigg[-\lambda\bigg
\{-\frac{f}{2}e^{\nu}+\big(\frac{\mu'}{r}-\frac{\mu'\nu'}{4}+\frac
{\mu''}{2}+\frac{\mu'^{2}}{4}\big)
\\\nonumber
&\times&\big(1-K
e^{\frac{-R}{B}}\big)+\big(\frac{\nu'}{2}-\frac{2}{r}\big)\big(\frac{1}{B}K
e^{\frac{-R}{B}}\big)R'-\big\{\big(\frac{1}{B}K
e^{\frac{-R}{B}}\big)R''
\\\nonumber
&+&\big(\frac{-1}{B^{2}}K
e^{\frac{-R}{B}}\big)R'^{2}\big\}\bigg\}+(3\lambda+2)
\bigg\{\frac{f}{2}e^{\nu}+\big(\frac{\mu''}{2}-\frac{\nu'}{r}-\frac
{\mu'\nu'}{4}+\frac{\mu'^{2}}{4}\big)
\\\nonumber
&\times&\big(1-K
e^{\frac{-R}{B}}\big)+\big(\frac{\mu'}{2}+\frac{2}{r}\big)\big(\frac{1}{B}K
e^{\frac{-R}{B}}\big)R'\bigg\}-2\lambda\bigg
\{\frac{f}{2}e^{\nu}-\big(\frac{(\mu'-\nu')r}{2}
\\\nonumber
&-&e^{\nu}+1\big) \frac{\big(1-K
e^{\frac{-R}{B}}\big)}{r^{2}}+\big(\frac{\mu'-\nu'}{2}+\frac{1}{r}\big)\big(\frac{1}{B}K
e^{\frac{-R}{B}}\big)R'+\big\{\big(\frac{1}{B}K
e^{\frac{-R}{B}}\big)
\\\label{A5}
&\times&R''+\big(\frac{-1}{B^{2}}K
e^{\frac{-R}{B}}\big)R'^{2}\big\}\bigg\}\bigg],
\\\nonumber
P_{t}&=&\frac{1}{{e^{\nu}}2(2\lambda+1)(\lambda+1)}\bigg[-\lambda\bigg
\{-\frac{f}{2}e^{\nu}+\big(\frac{\mu'}{r}-\frac{\mu'\nu'}
{4}+\frac{\mu''}{2}+\frac{\mu'^{2}}{4}\big)
\\\nonumber
&\times&\big(1-K
e^{\frac{-R}{B}}\big)+\big(\frac{\nu'}{2}-\frac{2}{r}\big)\big(\frac{1}{B}K
e^{\frac{-R}{B}}\big)R'-\big\{\big(\frac{1}{B}K
e^{\frac{-R}{B}}\big)R''
\\\nonumber
&+&\big(\frac{-1}{B^{2}}K
e^{\frac{-R}{B}}\big)R'^{2}\big\}\bigg\}+\lambda
\bigg\{\frac{f}{2}e^{\nu}+\big(\frac{\mu''}{2}-\frac{\nu'}{r}-\frac
{\mu'\nu'}{4}+\frac{\mu'^{2}}{4}\big)
\\\nonumber
&\times&\big(1-K e^{\frac{-R}{B}}\big)
+\big(\frac{\mu'}{2}+\frac{2}{r}\big)\big(\frac{1}{B}K
e^{\frac{-R}{B}}\big)R'\bigg\}-2(\lambda+1)\bigg\{\frac{f}{2}e^{\nu}
\\\nonumber
&-&\big(\frac{(\mu'-\nu')r}{2}-e^{\nu}+1\big)\frac{\big(1-K
e^{\frac{-R}{B}}\big)}{r^{2}}+\big
(\frac{\mu'-\nu'}{2}+\frac{1}{r}\big)\big(\frac{1}{B}K
e^{\frac{-R}{B}}\big)R'
\\\label{A6}
&+&\big\{\big(\frac{1}{B}K
e^{\frac{-R}{B}}\big)R''+\big(\frac{-1}{B^{2}}K
e^{\frac{-R}{B}}\big)R'^{2}\big\}\bigg\}\bigg].
\end{eqnarray}
The corresponding field equations corresponding to the model 2 are
\begin{eqnarray}\nonumber
\rho&=&\frac{1}{{e^{\nu}}2(2\lambda+1)(\lambda+1)}\bigg[(5\lambda+2)
\frac{1}{e^{\nu}}\bigg\{-\frac{f}{2}e^{\nu}+\big(\frac{\mu'}{r}-
\frac{\mu'\nu'}{4}+\frac{\mu''}{2}
\\\nonumber
&+&\frac{\mu'^{2}}{4}\big)\big(1-\frac{2nR\gamma
(1+\frac{R^{2}}{B^{2}})^{-1-n}}{B}\big)
+\big(\frac{\nu'}{2}-\frac{2}{r}\big)\bigg\{\frac{2n\gamma}{B(1+\frac
{R^{2}}{B^{2}})^{1+n}}
\\\nonumber
&\times&\bigg(\frac{(-2-2n)R^{2}}{B^{2}(1+\frac{R^{2}}{B^{2}})}-1\bigg)
R'\bigg\}-\bigg\{\frac{2n\gamma}{B(1+\frac{R^{2}}{B^{2}})^{1+n}}\bigg
(\frac{(-2-2n) R^{2}}{B^{2}(1+\frac{R^{2}}{B^{2}})}-1\bigg)
\\\nonumber
&\times&R''+\frac{4nR\gamma}{B^{3}(1+\frac{R^{2}}{B^{2}})^{2+n}}\bigg((3+
3n)-\frac{(8+6n+2n^{2})R^{2}}{B^{2}(1+\frac{R^{2}}{B^{2}})}\bigg)R'^{2}
\bigg\}\bigg\}
\\\nonumber
&+&\lambda
\bigg\{\frac{f}{2}e^{\nu}+\big(\frac{\mu''}{2}-\frac{\nu'}{r}-\frac
{\mu'\nu'}{4}+\frac{\mu'^{2}}{4}\big)\big(1-\frac{2nR\gamma(1+\frac
{R^{2}}{B^{2}})^{-1-n}}{B}\big)
\\\nonumber
&+&\big(\frac{\mu'}{2}+\frac{2}{r}\big)\bigg\{\frac{2n\gamma}{B(1
+\frac{R^{2}}{B^{2}})^{1+n}}\bigg(\frac{(-2-2n)R^{2}}{B^{2}(1+\frac
{R^{2}}{B^{2}})}-1\bigg)R'\bigg\}\bigg\}+2\lambda\bigg\{\frac{f}{2
}e^{\nu}
\\\nonumber
&+&\big(\frac{(\mu'-\nu')r}{2}-e^{\nu}+1\big)\frac{\big(-1+\frac
{2nR\gamma(1+\frac{R^{2}}{B^{2}})^{-1-n}}{B}\big)}{r^{2}}+\bigg(\frac
{\mu'-\nu'}{2}+\frac{1}{r}\big)
\\\nonumber
&\times&\bigg\{\frac{2n\gamma}{B(1+\frac{R^{2}}{B^{2}})^{1+n}}\bigg
(\frac{(-2-2n)R^{2}}{B^{2}(1+\frac{R^{2}}{B^{2}})}-1\bigg)R'\bigg\}
+\bigg\{\frac{2n\gamma}{B(1+\frac{R^{2}}{B^{2}})^{1+n}}
\\\nonumber
&\times&\bigg(\frac{(-2-2n)R^{2}}{B^{2}(1+\frac{R^{2}}{B^{2}})}-1\bigg)
R''+\frac{4nR\gamma}{B^{3}(1+\frac{R^{2}}{B^{2}})^{2+n}}
\bigg((3+3n)
\\\label{A7}
&-&\frac{(8+6n+2n^{2})R^{2}}{B^{2}(1+\frac{R^{2}}{B^{2}})}
\bigg)R'^{2}\bigg\}\bigg\}\bigg],
\\\nonumber
P_{r}&=&\frac{1}{{e^{\nu}}2(2\lambda+1)(\lambda+1)}\bigg[-\lambda\bigg
\{-\frac{f}{2}e^{\nu}+\big(\frac{\mu'}{r}-\frac{\mu'\nu'}{4}+\frac
{\mu''}{2}+\frac{\mu'^{2}}{4}\big)\big(1
\\\nonumber
&-&\frac{2nR\gamma
(1+\frac{R^{2}}{B^{2}})^{-1-n}}{B}\big)+\big(\frac{\nu'}{2}-\frac{2}{r}
\big)\bigg\{\frac{2n\gamma}{B(1+\frac{R^{2}}{B^{2}})^{1+n}}\bigg(\frac{
(-2-2n)R^{2}}{B^{2}(1+\frac{R^{2}}{B^{2}})}
\\\nonumber
&-&1\bigg)R'\bigg\}-\bigg\{\frac{2n\gamma}{B(1+\frac{R^{2}}{B^{2}})^{1+n}}
\bigg(\frac{(-2-2n)R^{2}}{B^{2}(1+\frac{R^{2}}{B^{2}})}-1\bigg)R''+\frac
{4nR\gamma}{B^{3}(1+\frac{R^{2}}{B^{2}})^{2+n}}
\\\nonumber
&\times&\bigg((3+3n)-\frac{(8+6n+2n^{2})R^{2}}{B^{2}(1+\frac{R^{2}}{B^{2}
})}\bigg)R'^{2}\bigg\}\bigg\}+(3\lambda+2)\bigg\{\frac{f}{2}e^{\nu}+\big
(\frac{\mu''}{2}-\frac{\nu'}{r}
\\\nonumber
&-&\frac{\mu'\nu'}
{4}+\frac{\mu'^{2}}{4}\big)\big(1-\frac{2nR\gamma(1+\frac{R^{2}}{B^{2}})
^{-1-n}}{B}\big)+\big(\frac{\mu'}{2}+\frac{2}{r}\big)\bigg\{\frac{2n\gamma}
{B(1+\frac{R^{2}}{B^{2}})^{1+n}}
\\\nonumber
&\times&\bigg(\frac{(-2-2n)R^{2}}{B^{2}(1+\frac{R^{2}}{B^{2}})}-1\bigg)
R'\bigg\}\bigg\}-2\lambda\bigg\{\frac{f}{2}e^{\nu}+\big(\frac{(\mu'-\nu')
r}{2}-e^{\nu}+1\big)
\\\nonumber
&\times&\frac{\big(-1+\frac{2nR\gamma(1+\frac{R^{2}}{B^{2}})^{-1-n}}{B}
\big)}{r^{2}}+\big(\frac{\mu'-\nu'}{2}+\frac{1}{r}\big)\bigg\{\frac
{2n\gamma}{B(1+\frac{R^{2}}{B^{2}})^{1+n}}
\\\nonumber
&\times&\bigg(\frac{(-2-2n)R^{2}}{B^{2}(1+\frac{R^{2}}{B^{2}})}-1\bigg)R'
\bigg\}+\bigg\{\frac{2n\gamma}{B(1+\frac{R^{2}}{B^{2}})^{1+n}}\bigg(\frac{
(-2-2n)R^{2}}{B^{2}(1+\frac{R^{2}}{B^{2}})}-1\bigg)R''
\\\label{A8}
&+&\frac{4nR\gamma}{B^{3}(1+\frac{R^{2}}{B^{2}})^{2+n}}\bigg((3+3n)-
\frac{(8+6n+2n^{2})R^{2}}{B^{2}(1+\frac{R^{2}}{B^{2}})}\bigg)R'^{2}
\bigg\}\bigg\}\bigg],
\\\nonumber
P_{t}&=&\frac{1}{{e^{\nu}}2(2\lambda+1)(\lambda+1)}\bigg[-\lambda
\bigg\{-\frac{f}{2}e^{\nu}+\big(\frac{\mu'}{r}-\frac{\mu'\nu'}
{4}+\frac{\mu''}{2}+\frac{\mu'^{2}}{4}\big)\big(1
\\\nonumber
&-&\frac{2nR\gamma
(1+\frac{R^{2}}{B^{2}})^{-1-n}}{B}\big)+\big(\frac{\nu'}{2}-\frac{2}
{r}\big)\bigg\{\frac{2n\gamma}{B(1+\frac{R^{2}}{B^{2}})^{1+n}}\bigg
(\frac{(-2-2n)R^{2}}{B^{2}(1+\frac{R^{2}}{B^{2}})}
\\\nonumber
&-&1\bigg)R'\bigg\}-\bigg\{\frac{2n\gamma}{B(1+\frac{R^{2}}{B^{2}}
)^{1+n}}\bigg(\frac{(-2-2n)R^{2}}{B^{2}(1+\frac{R^{2}}{B^{2}})}-1
\bigg)R''+\frac{4nR\gamma}{B^{3}(1+\frac{R^{2}}{B^{2}})^{2+n}}
\\\nonumber
&\times&\bigg((3+3n)-\frac{(8+6n+2n^{2})R^{2}}{B^{2}(1+\frac{R^{2}}
{B^{2}})}\bigg)R'^{2}\bigg\}\bigg\}+\lambda\bigg\{\frac{f}{2}e^{\nu}
+\big(\frac{\mu''}{2}-\frac{\nu'}{r}
\\\nonumber
&-&\frac{\mu'\nu'}{4}+\frac{\mu'^{2}}{4}\big)\big(1-\frac{2nR
\gamma(1+\frac{R^{2}}{B^{2}})^{-1-n}}{B}\big)+\big(\frac{\mu'}{2}
+\frac{2}{r}\big)\bigg\{\frac{2n\gamma}{B(1+\frac{R^{2}}{B^{2}})^{1+n}}
\\\nonumber
&\times&\bigg(\frac{(-2-2n)R^{2}}{B^{2}(1+\frac{R^{2}}{B^{2}})}-1\bigg)
R'\bigg\}\bigg\}-2(\lambda+1)\bigg\{\frac{f}{2}e^{\nu}+\big(\frac{
(\mu'-\nu')r}{2}-e^{\nu}+1\big)
\\\nonumber
&\times&\frac{\big(-1+\frac{2nR\gamma(1+\frac{R^{2}}{B^{2}})^{-1-n}}
{B}\big)}{r^{2}}+\big(\frac{\mu'-\nu'}{2}+\frac{1}{r}\big)\bigg
\{\frac{2n\gamma}{B(1+\frac{R^{2}}{B^{2}})^{1+n}}
\\\nonumber
&\times&\bigg(\frac{(-2-2n)R^{2}}{B^{2}(1+\frac{R^{2}}{B^{2}})}
-1\bigg)R'\bigg\}+\bigg\{\frac{2n\gamma}{B(1+\frac{R^{2}}{B^{2}})
^{1+n}}\bigg(\frac{(-2-2n)R^{2}}{B^{2}(1+\frac{R^{2}}{B^{2}})}
-1\bigg)R''
\\\label{A9}
&+&\frac{4nR\gamma}{B^{3}(1+\frac{R^{2}}{B^{2}})^{2+n}}\bigg(
(3+3n)-\frac{(8+6n+2n^{2})R^{2}}{B^{2}(1+\frac{R^{2}}{B^{2}})}
\bigg)R'^{2}\bigg\}\bigg\}\bigg].
\end{eqnarray}
The field equations for the model 3 take the following form
\begin{eqnarray}\nonumber
\rho&=&\frac{1}{{e^{\nu}}2(2\lambda+1)(\lambda+1)}\bigg[(5\lambda+2)
\bigg\{-\frac{f}{2}e^{\nu}+\big(\frac{\mu'}{r}-\frac{\mu'\nu'}
{4}+\frac{\mu''}{2}+\frac{\mu'^{2}}{4}\big)
\\\nonumber
&\times&\big(1-\omega\sec h^{2}\big(\frac{R}{B}\big)\big)+\big(\frac
{\nu'}{2}-\frac{2}{r}\big)\frac{2\omega\sec h^{2}(\frac{R}{B})\tanh
(\frac{R}{B})}{B}R'
\\\nonumber
&-&\bigg\{\frac{2\omega\sec h^{2}(\frac{R}{B})\tanh(\frac{R}{B})}
{B}R''+\frac{2\omega\sec h^{4}(\frac{R}{B})}{B^{2}}\big(1-2\sinh^
{2}\big(\frac{R}{B}\big)\big)R'^{2}\bigg\}\bigg\}
\\\nonumber
&+&\lambda\bigg\{\frac{f}{2}e^{\nu}+\big(\frac{\mu''}{2}-\frac
{\nu'}{r}-\frac{\mu'\nu'}{4}+\frac{\mu'^{2}}{4}\big)\big(1-
\omega\sec h^{2}\big(\frac{R}{B}\big)\big)+\big(\frac{\mu'}{2}
+\frac{2}{r}\big)
\\\nonumber
&\times&\frac{2\omega\sec h^{2}(\frac{R}{B})\tanh(\frac{R}{B})}
{B}R'\bigg\}+2\lambda\bigg\{\frac{f}{2}e^{\nu}+\big(\frac{
(\mu'-\nu')r}{2}-e^{\nu}+1\big)
\\\nonumber
&\times&\frac{\big(-1+\omega\sec h^{2}\big(\frac{R}{B}\big)\big)}
{r^{2}}+\big(\frac{\mu'-\nu'}{2}+\frac{1}{r}\big)\big(\frac{2
\omega\sec h^{2}(\frac{R}{B})\tanh(\frac{R}{B})}{B}\big)
\\\nonumber
&\times&R'+\bigg\{\frac{2\omega\sec h^{2}(\frac{R}{B})\tanh
(\frac{R}{B})}{B}R''+\frac{2\omega\sec h^{4}(\frac{R}{B})}{B^{2}}
\\\label{A10}
&\times&\big(1-2\sinh^{2}\big(\frac{R}{B}\big)\big)R'^{2}\bigg\}
\bigg\}\bigg],
\\\nonumber
P_{r}&=&\frac{1}{{e^{\nu}}2(2\lambda+1)(\lambda+1)}\bigg\{-\frac
{f}{2}e^{\nu}+\big(\frac{\mu'}{r}-\frac{\mu'\nu'}{4}+\frac
{\mu''}{2}+\frac{\mu'^{2}}{4}\big)\big(1
\\\nonumber
&-&\omega\sec h^{2}\big(\frac{R}{B}\big)\big)+\big(\frac{\nu'}{2}
-\frac{2}{r}\big)\frac{2\omega\sec h^{2}(\frac{R}{B})\tanh(\frac
{R}{B})}{B}R'
\\\nonumber
&-&\bigg\{\frac{2\omega\sec h^{2}(\frac{R}{B})\tanh(\frac{R}{B})}
{B}R''+\frac{2\omega\sec h^{4}(\frac{R}{B})}{B^{2}}\big(1-2\sinh^
{2}\big(\frac{R}{B}\big)\big)R'^{2}\bigg\}\bigg\}
\\\nonumber
&+&(3\lambda+2)\bigg\{\frac{f}{2}e^{\nu}+\big(\frac{\mu''}{2}
-\frac{\nu'}{r}-\frac{\mu'\nu'}{4}+\frac{\mu'^{2}}{4}\big)
\big(1-\omega\sec h^{2}\big(\frac{R}{B}\big)\big)
\\\nonumber
&+&\big(\frac{\mu'}{2}+\frac{2}{r}\big)\frac{2\omega\sec
h^{2}(\frac{R}{B})\tanh(\frac{R}{B})}{B}R'\bigg\}-2\lambda
\bigg\{\frac{f}{2}e^{\nu}+\big(\frac{(\mu'-\nu')r}{2}
\\\nonumber
&-&e^{\nu}+1\big)\frac{\big(-1+\omega\sec h^{2}\big
(\frac{R}{B}\big)\big)}{r^{2}}+\big(\frac{\mu'-\nu'}{2}
+\frac{1}{r}\big)\big(\frac{2\omega\sec h^{2}(\frac{R}{B})
\tanh(\frac{R}{B})}{B}\big)R'
\\\nonumber
&+&\bigg\{\frac{2\omega\sec h^{2}(\frac{R}{B})\tanh(\frac{R}{B})}
{B}R''
\\\label{A11}
&+&\frac{2\omega\sec h^{4}(\frac{R}{B})}{B^{2}}\big(1-2\sinh^
{2}\big(\frac{R}{B}\big)\big)R'^{2}\bigg\}\bigg\}\bigg],
\\\nonumber
P_{t}&=&\frac{1}{{e^{\nu}}2(2\lambda+1)(\lambda+1)}\bigg[-\lambda
\bigg\{-\frac{f}{2}e^{\nu}+\big(\frac{\mu'}{r}-\frac{\mu'\nu'}
{4}+\frac{\mu''}{2}+\frac{\mu'^{2}}{4}\big)
\\\nonumber
&\times&\big(1-\omega\sec h^{2}\big(\frac{R}{B}\big)\big)+\big(\frac
{\nu'}{2}-\frac{2}{r}\big)\frac{2\omega\sec h^{2}(\frac{R}{B})\tanh
(\frac{R}{B})}{B}R'
\\\nonumber
&-&\bigg\{\frac{2\omega\sec h^{2}(\frac{R}{B})\tanh(\frac{R}{B})}
{B}R''+\frac{2\omega\sec h^{4}(\frac{R}{B})}{B^{2}}\big(1-2\sinh^{2}
\big(\frac{R}{B}\big)\big)R'^{2}\bigg\}\bigg\}
\\\nonumber
&+&\lambda\bigg\{\frac{f}{2}e^{\nu}+\big(\frac{\mu''}{2}-\frac
{\nu'}{r}-\frac{\mu'\nu'}{4}+\frac{\mu'^{2}}{4}\big)\big(1-\omega\sec
h^{2}\big(\frac{R}{B}\big)\big)+\big(\frac{\mu'}{2}+\frac{2}{r}\big)
\\\nonumber
&\times&\frac{2\omega\sec h^{2}(\frac{R}{B})\tanh(\frac{R}{B})}{B}R'
\bigg\}-2(\lambda+1)\bigg\{\frac{f}{2}e^{\nu}+\big(\frac{(\mu'-\nu')
r}{2}-e^{\nu}+1\big)
\\\nonumber
&\times&\frac{\big(-1+\omega\sec h^{2}\big(\frac{R}{B}\big)\big)}
{r^{2}}+\big(\frac{\mu'-\nu'}{2}+\frac{1}{r}\big)\big(\frac{2\omega\sec
h^{2}(\frac{R}{B})\tanh(\frac{R}{B})}{B}\big)R'
\\\nonumber
&+&\bigg\{\frac{2R''\omega\sec h^{2}(\frac{R}{B})\tanh(\frac{R}{B})}
{B}
\\\label{A12}
&+&\frac{2\omega\sec h^{4}(\frac{R}{B})\big(1-2\sinh^{2}
\big(\frac{R}{B}\big)\big)R'^{2}}{B^{2}}\bigg\}\bigg\}\bigg].
\end{eqnarray}
\\
\textbf{Acknowledgement:} We are thankful to Dr. Zeeshan Gul for
fruitful discussions during the write up of this paper.\\\\
\textbf{Data Availability Statement:} This manuscript has no
associated data.

\end{document}